\documentclass[twocolumn,prd,showpacs]
              {revtex4}

\usepackage{graphicx,color}

\begin{document}

\title{Modified Friedman scenario from the Wheeler-DeWitt equation}

\author{Michael~Maziashvili}
\email{maziashvili@ictsu.tsu.edu.ge} \affiliation{Department of
Theoretical Physics, Tbilisi State University, \\ 3 Chavchavadze
Ave., Tbilisi 0128, Georgia}

\begin{abstract}
We consider the possible modification of the Friedman equation
  due to operator ordering parameter entering the Wheeler-DeWitt
  equation.
\end{abstract}
% 04.60.Kz Lower dimensional models; minisuperspace models.
% 98.80.Qc Quantum cosmology.

\maketitle

The standard approach to quantum cosmology consists of quantizing
a minisuperspace model. The corresponding Wheeler-DeWitt (WDW)
equation \cite{WDW} contains an arbitrary operator ordering
parameter. In one of our papers \cite{Ma} it was noticed that the
operator ordering term for large values of operator ordering
parameter can affect the matching of false vacuum instanton with
the Coleman-De Luccia bounce. Continuing in the spirit of that
paper, in this brief note we would like to consider how the
operator ordering term for large enough values of operator
ordering parameter can affect the Friedman equation considered as
a semiclassical limit of the WDW equation. (Throughout this paper
we assume $c=1$).

Let us consider a closed universe filled with a vacuum of constant
energy density and the radiation,
\[\rho(a)=\rho_v+\frac{\epsilon}{a^4}~,\] where $\rho_v$ is the
vacuum energy density, $a$ is the scale factor and $\epsilon$ is a
constant characterizing the amount of radiation. The evolution
equation for $a$ can be written as
\begin{equation}\label{Fri}-\frac{3\pi}{4G}a\dot{a}^2-\frac{3\pi}{4G}a+2\pi^2a^3\rho(a)=0~.\end{equation}
This equation is identical to that for a particle moving in a
potential \[U(a)=\frac{3\pi}{4G}-2\pi^2a^2\rho(a)~,\] with zero
energy, Fig.1.

\begin{figure}
  \begin{center}

\includegraphics{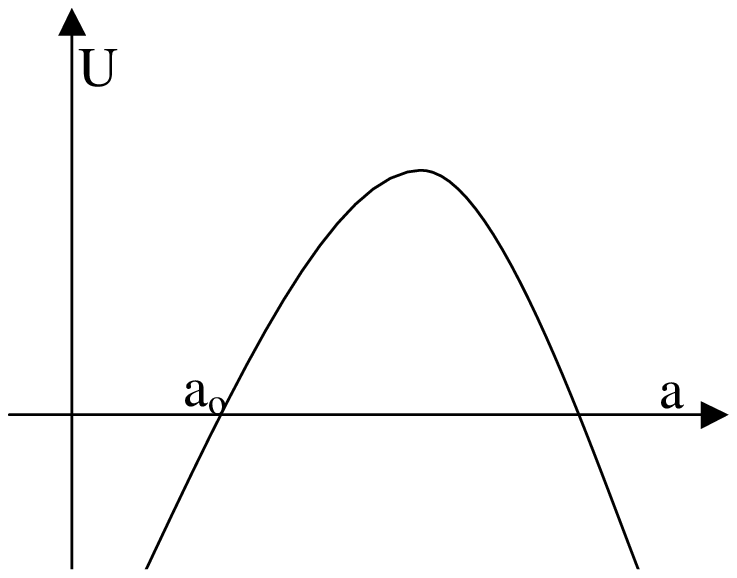}

 \end{center}

  \caption{A schematic picture of the potential $U(a)$.}

\end{figure}

We tacitely assumed $256\,\pi^2G^2\rho_v\,\epsilon/9<1~.$ The
universe can start at $a=0$, expand to a maximum radius $a_0$ and
then recollapse or tunnel through the potential barrier to the
regime of unbounded expansion. Eq.(\ref{Fri}) represents the zero
energy condition for the Lagrangian
\begin{equation}{\cal L}=-\frac{3\pi}{4G}a\dot{a}^2+\frac{3\pi a}{4G}-2\pi^2a^3\rho(a)~.\end{equation}
To quantize the model, we find the Hamiltonian
\begin{equation}\label{Ham}{\cal H}=-\frac{G}{3\pi}\frac{p^2}{a}-\frac{3\pi}{4G}a+2\pi^2a^3\rho(a)~,\end{equation}
and replace the momentum in (\ref{Ham}) by a differential operator
$p\rightarrow -i\hbar d/da$. Then the WDW equation is written down
as
\begin{equation}\label{WDW}\left(\frac{G\hbar^2}{3\pi}\frac{d^2}{da^2}+\frac{Gq\hbar^2}{3\pi a}\frac{d}{da}-\frac{3\pi}{4G}a^2+2\pi^2a^4\rho(a)\right)\psi(a)=0~,\end{equation}
where $q$ is an arbitrary parameter corresponding to the operator
ordering $p^2/a=a^{-(q+1)}p\,a^qp$. Throughout this paper $q$
denotes \[q\equiv\alpha\frac{a_0^2}{\hbar\,G}~,\] where $\alpha$
is an arbitrary dimensionless parameter. Disregarding the operator
ordering term, the eq.(\ref{WDW}) has the form of a
one-dimensional Schr\"{o}dinger equation for a particle described
by a coordinate $a$, having zero energy and moving in the
potential
\[V(a)=\frac{3\pi}{4G}a^2-2\pi^2a^4\rho(a)~.\]
Assuming that $\alpha$ is sufficiently large, in the WKB
approximation \cite{LLQ}

\[\psi(a)=\exp\left\{-iW_0(a)/\hbar+W_1+\cdots\right\}~,\]

to the lowest order in $\hbar$ one obtains

\begin{equation}\label{HJ}-\frac{G}{3\pi}\left(\frac{dW_0}{da}\right)^2-\frac{iGq\hbar}{3\pi a}\frac{dW_0}{da}-\frac{3\pi a^2}{4G}+2\pi^2a^4\rho(a)=0~.\end{equation}
We have assumed that for large values of $\alpha$ the second term
in eq.(\ref{HJ}) may be of the order of the first one. Let us
judge the validity of this assumption. From eq.(\ref{HJ}) one
obtains

\begin{equation}\label{sol}\frac{dW_0}{da}=-i\frac{q\hbar}{2a}\pm i\frac{\sqrt{f(a)}}{2}~,\end{equation}

where
\[f(a)=\frac{q^2\hbar^2}{a^2}+\frac{9\pi^2a^2}{G^2}-\frac{24\pi^3a^4\rho(a)}{G}~.\]

In what follows, in the region $a<a_0$, we assume the following
boundary condition
\begin{equation}\label{bc}\frac{dW_0}{da}=-i\frac{q\hbar}{2a}-i\frac{\sqrt{f(a)}}{2}~.\end{equation}
Under this assumption, in the region $a<a_0$, for large enough
values of $\alpha$ the ratio of the first and second terms in
eq.(\ref{WDW}) takes the form
\begin{equation}\label{vcon}\left|\frac{a}{q\hbar}\frac{dW_0}{da}\right|=\left|\frac{1}{2}+\frac{1}{2}\sqrt{1+\frac{9\pi^2a^4}{\alpha^2 a_0^4}-\frac{24G\pi^3a^6\rho(a)}{\alpha^2 a_0^4}} \,\right|\sim 1~.\end{equation}
So, under the boundary condition (\ref{bc}), in the region
$a<a_0$, the second term in eq.(\ref{WDW}) for large values of
$\alpha$ can be of the order of the first one and must be kept.
Notice that for the solution (\ref{sol}) with the positive sign
the second term in eq.(\ref{HJ}) becomes suppressed in comparison
with the first one and should be omitted.

Let us now check the standard WKB validity condition \cite{LLQ}.
For large values of $\alpha$ in the region $a<a_0$ one obtaines
\[\frac{dW_0}{da}\sim -i\frac{a_0^2\,\alpha }{G\,a}~,~~\frac{d^2W_0}{da^2}\sim
i\frac{a_0^2\,\alpha}{G\,a^2}~,\] and correspondingly
\begin{equation}\label{sWKBc}\hbar\left|\frac{d^2W_0}{da^2}\left/\left(\frac{dW_0}{da}\right.\right)^2\right|\sim q^{-1}\ll 1~.\end{equation}
Thus, for the solution (\ref{bc}) the approximation (\ref{HJ}) is
justified in the region $a<a_0$ for large values of $\alpha$.

From the Hamilton-Jacobi equation (\ref{HJ}) one obtaines the
following Lagrangian \cite{LLM}
\[{\cal L}=-\frac{3\pi}{4G}a\dot{a}^2-\frac{iq\hbar}{2}\frac{\dot{a}}{a}
+\frac{Gq^2\hbar^2}{12\pi a^3}+\frac{3\pi
a}{4G}-2\pi^2a^3\rho(a)~.\] In this Lagrangian one can omit the
second term because it contains the total time derivative and
thereby does not affect the equation of motion \cite{LLM}.
Therefore one gets

\[{\cal
L}=-\frac{3\pi}{4G}a\dot{a}^2+\frac{Gq^2\hbar^2}{12\pi
a^3}+\frac{3\pi a}{4G}-2\pi^2a^3\rho(a)~,\]

for which the zero-energy condition takes the form

\begin{equation}\label{mFri}-\frac{3\pi}{4G}a\dot{a}^2-\frac{\alpha^2a_0^4}{12G\pi a^3}-
\frac{3\pi}{4G}a+2\pi^2a^3\rho(a)=0~.\end{equation}

The modified Friedman equation (\ref{mFri}) containes a new term,
which decays very fast in the course of expansion of the universe
and therefore can not affect the late time cosmology. But,
however, the appearance of this new term is quite important for it
can avoid the collapse of the universe.

To summarize, we considered the influence of the operator ordering
term on the semiclassical limit of the WDW equation for large
enough values of operator ordering parameter. More precisely from
the very beginning in eq.(\ref{WDW}) we assume $q=\alpha
a_0^2/\hbar G$, where $\alpha$ is an arbitrary dimensionless
parameter and take the boundary condition (\ref{bc}) in the region
$a<a_0.$ For large values of $\alpha$, such that to ensure
validity of eqs.(\ref{vcon},\,\,\ref{sWKBc}), the semiclassical
limit of the WDW equation produces a modification of the Friedman
equation eq.(\ref{mFri}). The new term appearing in equation
(\ref{mFri}) decays very fast and therefore does not affect the
cosmology in the course of expansion of the universe, but,
however, can avoid the collapse of the universe.

\begin{acknowledgments}
The author is greatly indebted to the Abdus Salam International
Center for Theoretical Physics, where this paper was begun, for
hospitality. It is a pleasure to acknowledge helpful conversations
with Professors A.~Dolgov, M.~Gogberashvili and V.~Rubakov.
\end{acknowledgments}

\end{document}